\documentclass[aps,prl,twocolumn]{revtex4}
\usepackage{graphicx}

\begin{document}

\title{Spin-injection through an Fe/InAs Interface}
\author{M. Zwierzycki$^{1*}$, K. Xia$^{1+}$, P. J. Kelly$^1$, 
G. E. W. Bauer$^2$, and I. Turek$^3$}

\affiliation{$^1$Faculty of Applied Physics and MESA$^{+}$ Research
Institute, 
University of Twente, P.O. Box 217, 7500 AE Enschede, The Netherlands\\
$^2$Department of Applied Physics and DIMES, Delft University of Technology, 
Lorentzweg 1, 2628 CJ Delft, The Netherlands \\
$^3$Institute of Physics of Materials, Academy of Sciences
of the Czech Republic, CZ-616 62 Brno, Czech Republic}
\date{\today}

\begin{abstract}
 The spin-dependence of the interface resistance between ferromagnetic
 Fe and InAs is calculated from first-principles for specular and 
 disordered (001) interfaces. Because of the symmetry mismatch in the 
 minority-spin channel, the specular interface acts as an efficient
 spin filter with a transmitted current polarisation between 98 an 89\%.
 The resistance of a specular interface in the diffusive regime is 
 comparable to the resistance of a few microns of bulk InAs.
 Symmetry-breaking arising from interface disorder reduces the spin 
 asymmetry substantially and we conclude that efficient spin injection
 from Fe into InAs can only be realized using high quality epitaxial
 interfaces.
\end{abstract}
\maketitle

Compared to magnetic multilayers, semiconductor heterostructures can
be made with low defect concentrations, resulting in large transport
mean-free paths at low temperatures. The large Fermi wavelength of
charge carriers in doped semiconductors then leads to finite-size 
effects and a host of phenomena related to the electron waves 
maintaining their phase coherence over long distances \cite{SSP91}. 
Current interest in spin-injection into semiconductors is motivated
by a desire to combine the control over transport phenomena possible
in semiconductors using external gates with the additional spin degree
of freedom in ferromagnetic metals which has given rise to such new 
phenomena as oscillatory exchange coupling, giant magnetoresistance
(GMR), and junction (or tunnel) magneto-resistance (JMR or TMR)
\cite{GMR_reviews}. 

While spin-injection from a magnetic semiconductor using optical
detection techniques was successfully demonstrated some years ago
\cite{Fiederling99,Ohno99}, spin-injection from a metallic ferromagnet
into a semiconductor was only realized very recently \cite{Zhu01}.
Schmidt {\em et al.} \cite{Schmidt00} pointed out that a basic
obstacle to spin-injection in this case is the large difference in
their conductivities; the spin-independent resistivity of a
semiconductor such as InAs is much larger than either the majority- or
minority-spin resistivity of a ferromagnetic metal (FM) such as Fe.
The resistances added in series are dominated by the spin-independent
semiconductor term. Schmidt \emph{et al.} did not take into account the
possibility of a spin-dependent interface resistance which, if
sufficiently large, could generate a spin-dependent potential drop at
the interface \cite{Rashba00,Fert01}. 

Qualitative arguments have been given for the existence of such a
spin-dependence \cite{Kirczenow01} and a number of studies based on
free-electron models have appeared \cite{Grundler01,Hu01}. Transition
metal atoms are characterized by five-fold orbitally degenerate $d$
states with a large Hund's rule exchange splitting leading to large 
spin magnetic moments. In a solid these 10 $d$ states form complex 
band structures and Fermi surfaces. The origin of the spin-dependence
of the interface resistance in magnetic multilayers lies in the
difference between how the majority- and minority-spin states match to
the spin-degenerate electron states in a non-magnetic metal (NM)
\cite{Schep9598,Zahn95}. By expressing the (mis)matching at the FM/NM
interface in terms of the reflection and transmission matrices of
scattering theory \cite{Datta95}, the corresponding resistances can be
calculated within the framework of the Landauer-B\"uttiker transport
formalism \cite{Schep97,Stiles00,Xia01}. Because free-electron models
do not describe realistically the electronic structure and magnetism 
of transition metal elements and their interfaces with other materials,
we have calculated the spin-dependent transmission for the Fe/InAs(001) 
system including the full electronic band structures and derive the 
corresponding interface resistances. We argue that spin-injection 
should be observable for specular interfaces. However, we also show 
that introducing disorder into the interface layer greatly reduces the
spin asymmetry. We focus on the special case of Fe/InAs because it 
forms an Ohmic contact. Since there is no Schottky barrier between 
these two materials, it should be possible to realize much larger 
currents than for systems such as Fe/GaAs where electrons must tunnel 
through a Schottky barrier \cite{Zhu01}.

\begin{figure}
\centering
\resizebox{0.45\textwidth}{!}{\includegraphics[clip=true]{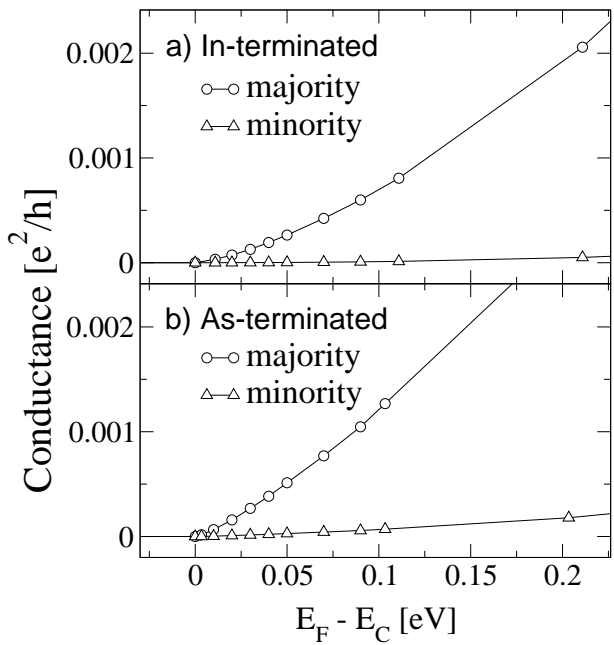}}
\caption{ Conductance of a) In- and b) As-terminated Fe/InAs
  interfaces (in units of $ e^2/h $) as a function of the distance of 
  the Fermi energy ($E_F$) from the bottom of the conductance band ($E_C$) 
  for majority ({\large $\circ $}) and  minority ($\triangle $) spins.}
\label{Fig1}
\end{figure}
To calculate transmission and reflection matrices without introducing
arbitrary fitting parameters, we use the local-density approximation
(LDA) of density functional theory. The present implementation is based
on the surface Green's function method \cite{Turek} formulated for
tight-binding linear muffin tin orbitals (TB-LMTO) \cite{Andersen85}.
Because a minimal basis set is used, we are able to model the 
disorder with lateral supercells. The calculations were carried out
in two stages. First the electronic structure, spin-densities and
potentials of Fe/InAs were determined self-consistently using the
layered TB-LMTO surface Green's function method \cite{Turek}. To take
account of the 5\% lattice mismatch between the bulk crystal structures
($a_{\mathrm{Fe}}=2.866$\AA, $a_{\mathrm{InAs}}=6.058$\AA), we assume Fe
to be tetragonal with its in-plane lattice constant matched to that of
the cubic InAs substrate. The vertical lattice constant is contracted 
so that the Fe unit cell volume is the same as for the bulk material.
To achieve reasonable space filling for InAs in the atomic spheres 
approximation \cite{Andersen85}, empty spheres were introduced in the 
interstitial positions in both In and As layers. The correct band gap 
for InAs (which is found to be metallic in a straightforward LDA 
calculation) was obtained using a ``scissors-operator'' correction 
term. To form an Ohmic contact, the Fermi level was positioned at the 
bottom of the InAs conduction band by using the CPA approximation to 
substitute some of the trivalent In with quadrivalent Sn which acts 
as an electron donor. The atomic potentials were calculated 
self-consistently for the 4 monolayers of Fe and 6 layers of InAs
closest to the interface \cite{InAs_pot}. The CPA was also used to 
determine self-consistent potentials for disordered interfaces 
\cite{Turek}. Using these potentials as input, the transmission 
coefficients $t^{\sigma}_{\mu\nu}(\mathbf{k}_{||})$, ($\mu$ and $\nu$
denote the incoming and transmitted Bloch waves, respectively) were
calculated in a second step with a recently developed scheme based on
the TB-LMTO method \cite{Xia01}. To calculate the conductance, a 
summation must be carried out over the two-dimensional (2D) Brillouin
zone (BZ). This was done with a ${\bf k_{\parallel}}$ mesh density 
equivalent to $6.4\times 10^5$ mesh points in the 2D BZ of a 
$1\times 1$ interface unit cell.

In Figure~\ref{Fig1} the spin-dependent conductances for In- and
As-terminated specular ($\mathbf{k}_{||}$ conserving) interfaces,
$G_{\sigma}=\sum_{\mu,\nu,k_{||}} |t_{\mu\nu}(\mathbf{k}_{||})|^2$,
are shown as a function of the position of the Fermi energy ($E_F$)
above the bottom of the conduction band ($E_C$) which is controlled
by the doping in an experiment \cite{InAs_pot}. For both terminations
a large spin-asymmetry is predicted.  For 
$E_F - E_C = 0.02\:\mathrm{eV}$
(corresponding to a doping concentration of about 
$10^{17}\mathrm{cm}^{-3}$, thus just in the metallic regime) the 
ratio $G_\uparrow / G_\downarrow $ is about 110 for In- and 18 for 
As-termination and decreases slowly with increasing $E_F-E_C$.  
These ratios correspond to current polarisation values (defined 
as $G_\uparrow - G_\downarrow / G_\uparrow + G_\downarrow $) of 
98 an 89\%, respectively.

\begin{figure}
 \resizebox{0.48\textwidth}{!}{\includegraphics{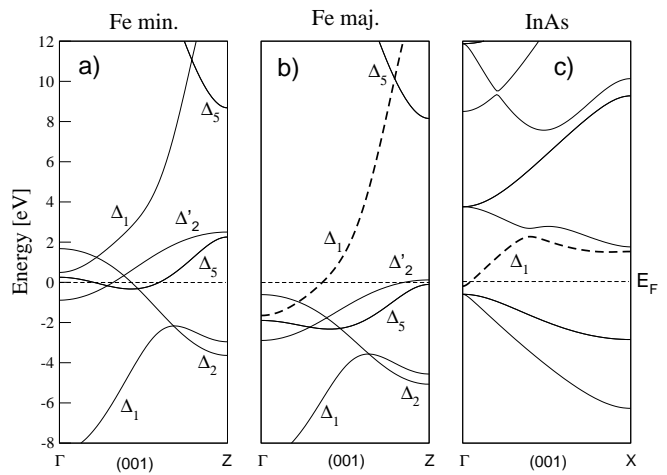}}    
  \caption{ Energy band structures of 
    tetragonal Fe minority spin states (a), majority spin states (b),
    and InAs states (c) at ${\bf k_{\parallel}}=0$ for ${\bf k} = (0 0
    k_z)$ perpendicular to the interface. The states with $\Delta_1$
    symmetry discussed in the text are shown dashed.}
\label{Fig2}
\end{figure}
\begin{figure}
 \begin{center}
   \resizebox{0.4\textwidth}{!}{
     \includegraphics[clip=false,draft=false]{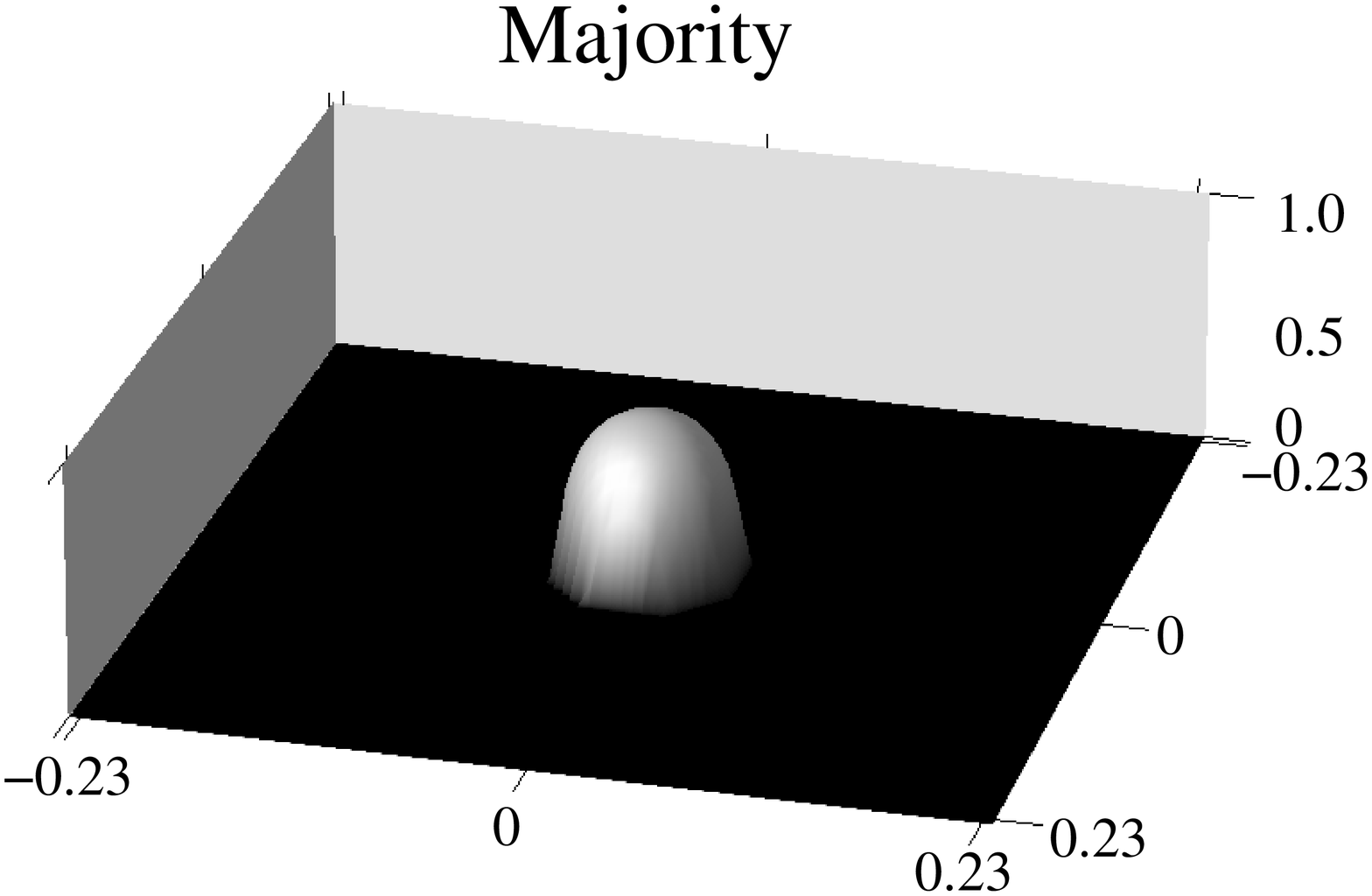}}\\
   \resizebox{0.4\textwidth}{!}{
     \includegraphics[clip=false,draft=false]{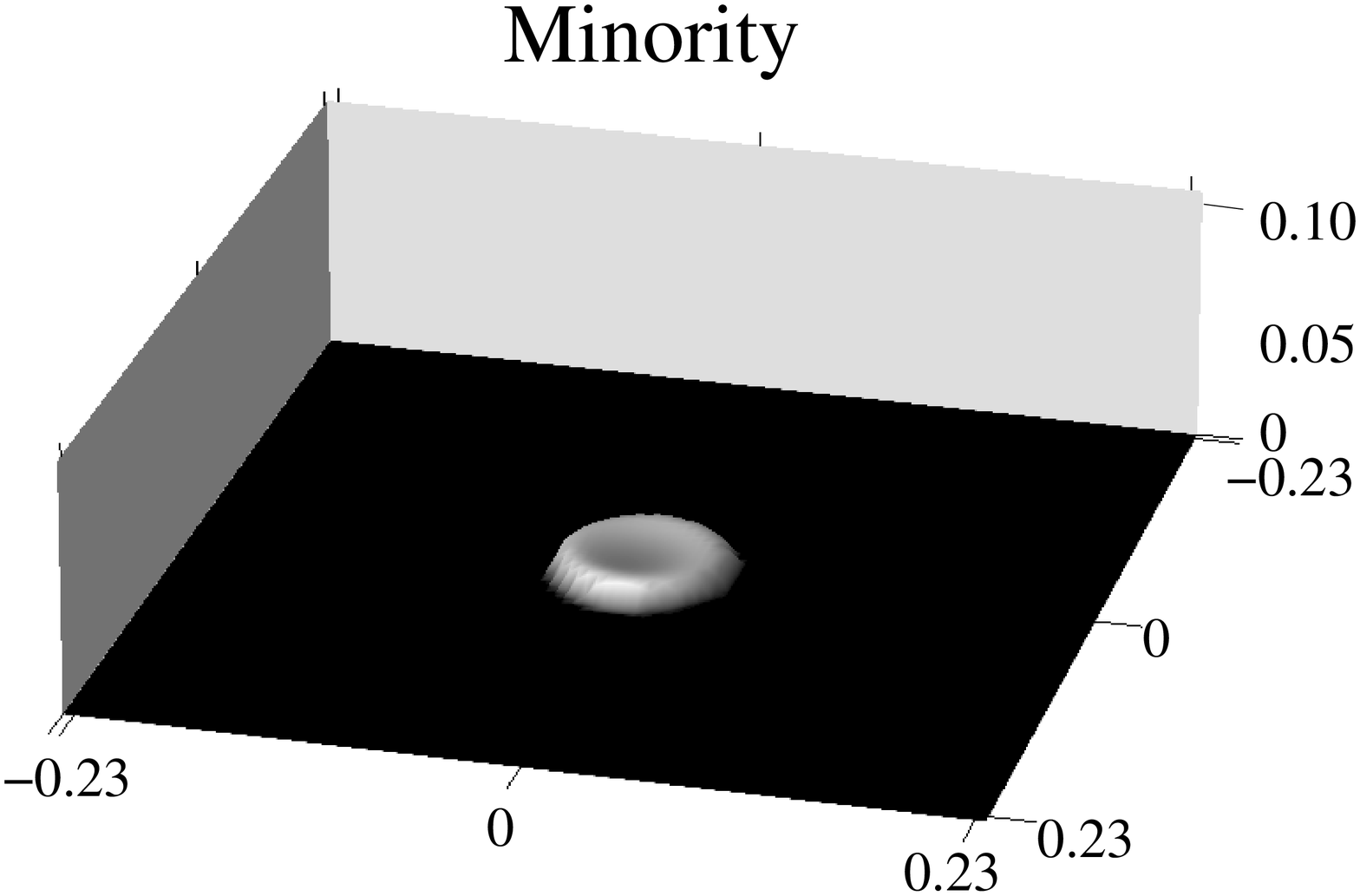}}
 \end{center}
\caption{The transmission probabilities as a function of ${\bf
    k}_{||}$ for the In-terminated interface (the As-terminated case is
  qualitatively similar). The upper plot is for majority- and the lower
  for minority-spins. $E_F - E_C$ is equal to 0.1 and only the central
  5\% of the 2DBZ area is shown (the values on the horizontal axes are
  given in the units of $\sqrt{2}\pi/a_{\mathrm{InAs}}$).  Note that
  different vertical scales are used for majority and minority spins.}
\label{Fig3}
\end{figure}

The large spin-dependence of the Fe/InAs interface scattering can be
simply understood by considering the bulk band structures of both
materials, shown in Fig.~\ref{Fig2} for ${\bf k_{\parallel}}=0$
($\bar{\Gamma}$ point).  We first note that the only states available
for transport in InAs are concentrated around the centre of the 2D BZ.
The ${\bf k_{\parallel}}$-resolved transmission coefficients are
therefore non-zero only close to the zone centre.  At ${\bf
  k_{\parallel}}=0$, the single occupied InAs conduction band state
has $\Delta_1$ symmetry. Comparing the Fe majority- and minority-spin
band structure, we immediately notice that only the majority bands
have a state with this symmetry at (or close to) the Fermi level.
Because the point group of the Fe/InAs(001) interface does not contain
a 4-fold rotation axis, the $\Delta'_2$ Fe states can also couple to
the $\Delta_1$ states in InAs. However, the $\Delta'_2$ states
consist of localized in-plane $d_{xy}$ orbitals so this coupling is
expected to be much smaller than between $\Delta_1$ states.  Though
this symmetry argument is only strictly applicable at ${\bf
  k_{\parallel}}=0$, the majority channel is expected to dominate the
conductance. The qualitative predictions are confirmed by the full
calculation. For $E_F - E_C = 0.1\:\mathrm{eV}$, the transmission
probability is plotted in Fig.~\ref{Fig3} as a function of
$\mathbf{k}_{||}$ for an In-terminated interface. For majority spins,
it has a maximum value $\sim 0.64$ at the $\bar{\Gamma}$ point. For
the minority spins it is a local minimum with a value almost two
orders of magnitude smaller.

The above results are only applicable in the ballistic regime where 
the resistance is dominated by interface scattering. To address the
more realistic diffusive regime, we need to determine the interface
resistance encountered when a specular interface is embedded in 
diffusive Fe and InAs. To do this, we use the expression derived by
Schep {\em et al.}  \cite{Schep97}:
\begin{displaymath}
  R_{\mathrm{Fe}/\mathrm{InAs}}=\frac{h}{e^2}
  \left[ 
    \frac{1}{\sum |t_{\mu\nu}|^2}
        -\frac{1}{2}\left(
                      \frac{1}{N_{\mathtt{Fe}}} + \frac{1}{N_{\mathrm{InAs}}}
                    \right)
  \right]
\end{displaymath}
where the first term is the inverse of the Landauer-B\"uttiker
conductance and $N_{\mathrm{Fe(InAs)}}$ is the Sharvin conductance
(in units of $ e^2/h $) of Fe (InAs). Obviously, the Sharvin 
resistance correction is dominated by the contribution from the 
semiconductor. 

For $E_F - E_C = 0.02\:\mathrm{eV}$ we obtain values of
$R^{\mathrm maj}=5.5 \times 10^4$ and 
$R^{\mathrm min}=7.1 \times 10^6\: \mathrm{f}\Omega \mathrm{m}^2$ 
for In-termination and 
$2.1 \times 10^4$, respectively, 
$5.2 \times 10^5 \mathrm{f}\Omega \mathrm{m}^2$ for As-termination.
All these resistances are much larger than the resistances seen by
electrons originating in the ferromagnetic layer within a spin-flip
diffusion length of the interface \cite{Schmidt00} which underlines 
the importance of interface properties (and not the bulk polarisation)
for injecting spins. Spin-injection can occur in diffusive systems
when the interface resistance is spin-dependent and comparable to the
resistance of the semiconductor layer \cite{Rashba00,Fert01}. Within
the free-electron model (with $m_{\mathrm{eff}}=0.04$) and assuming
Thomas-Fermi screening of the impurity potential \cite{Abram} we 
estimate the low-temperature resistivity of InAs with doping of
$10^{17}\mathrm{cm}^{-3}$ to be 
$\rho_{\mathrm InAs}=0.3 \times 10^{-4}\Omega\mathrm{m}$. 
A thickness $L$ of InAs has a resistance comparable to the interface
resistance $R_{\mathrm{Fe/InAs}}$ when 
$L \sim R_{\mathrm{Fe/InAs}}/\rho_{\mathrm InAs}$ which yields values 
ranging from $0.7 \mu m$ (majority spin, As termination) to 
$240 \mu m$ (minority spin, In termination).  

\begin{figure}
\begin{center}
  \resizebox{0.4\textwidth}{!}{\includegraphics[clip=true]{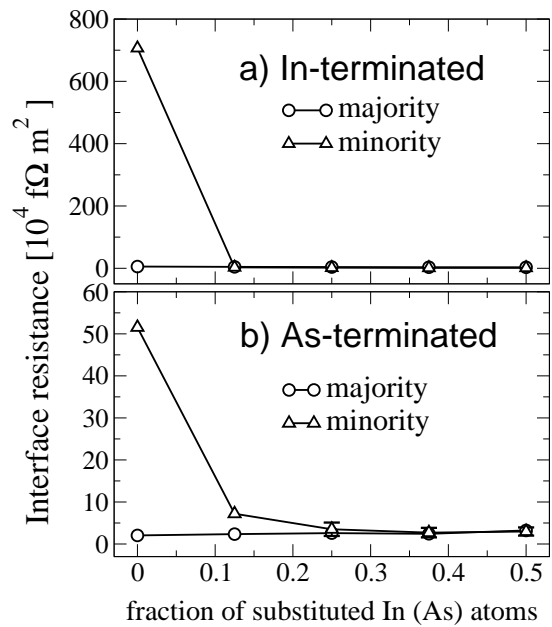}}
\end{center}
\caption{Interface resistances for a) In- and b) As-termination as a
  function of the fraction of interfacial In or As atoms substituted
  by Fe for majority ({\large $\circ $}) and minority ($\triangle $)
  spins.  For both terminations the symmetry-induced spin-asymmetry is
  strongly reduced by disorder.}
\label{Fig4}
\end{figure}

The large spin-dependence of specular interface scattering discussed
so far results directly from the symmetry of the Bloch states on
either side of the interface. It is important to know whether the
corresponding spin asymmetry will survive the interface disorder which
is invariably present at real interfaces.  To address this question we
performed calculations with $2 \times 2$ lateral supercells containing
8 In (As) atoms and 16 Fe atoms and introduced symmetry-breaking by
randomly replacing some of the interfacial In(As) atoms with iron.
Figure~\ref{Fig4} shows the majority- and minority-spin interface
resistances as a function of the fraction of In (As) atoms which were
replaced. The spread of values obtained for different configurations
is indicated, where applicable, by vertical error bars. For both
terminations we see relatively weak variation in the majority channel.
However, the large values of minority-spin interface resistances are
suppressed by interface disorder and soon assume values comparable to
the majority-spin values. This result suggests that the realization of
the strong spin-filtering effect predicted in our
calculations for In-terminated specular interfaces (and independently
by Wunnicke {\em et al.} for Fe/GaAs and Fe/ZnSe \cite{Wunnicke02})
requires very considerable care in preparing the interfaces. Since our
findings are based on symmetry arguments, they should be equally
applicable to epitaxial Fe/GaAs \cite{Zhu01} or Fe/AlGaAs
\cite{Hanbicki02} if the Schottky barrier is sufficiently thin that
carrier injection does not occur by thermionic emission over the
barrier. When tunnel barriers composed of amorphous oxides are used
for spin injection \cite{Motsnyi}, the symmetry arguments are most
likely no longer valid. Such systems need to be studied in more
detail.

The finding that disorder strongly {\em reduces} the high resistance 
of the minority-spin electrons by opening new transport channels which
are symmetry forbidden for specular interfaces is quite similar to what
was found for the Fe/Cr system by Xia {\em et al.} \cite{Xia01}.

In conclusion, we have studied the transport properties of Fe/InAs(001)
interfaces, taking into account the full electronic structure of both 
materials  and found strong spin selectivity of the transport through
these interfaces provided that they are grown epitaxially with a very
high degree of perfection.

This work is part of the research program for the ``Stichting voor
Fundamenteel Onderzoek der Materie'' (FOM) and was supported by 
the RT Network ``Computational Magnetoelectronics'' 
(contract No. RTN1-1999-00145),
the NEDO International Joint Research Grant Program 
``Nano-magnetoelectronics'', as well as by 
the Grant Agency of the Czech Republic (202/00/0122).

\noindent
$^*$ Permanent address: Institute of Molecular Physics,
P.A.N., Smoluchowskiego 17, 60-179 Poznan, Poland.\\
$^+$ Present address: Department of Physics, 
North Carolina State University, Raleigh, NC 27695, USA.


\begin{thebibliography}{}
\bibitem{InAs_pot} For simplicity, the self consistent calculation was
only carried out for one doping concentration (in the metallic regime).
Other doping concentrations were modelled by adding an appropriate 
constant potential shift on the InAs side of the interface.
\end{thebibliography}
\end{document}